\documentclass{article}
\usepackage[dvips]{graphicx}

\def\be{\begin{equation}}
\def\ee{\end{equation}}
\def\bea{\begin{eqnarray}}
\def\eea{\end{eqnarray}}
\def\ba{\begin{array}}
\def\ea{\end{array}}
\def\la{\label}
\def\rf{\ref}

\begin{document}
\title{The Application of the Yang-Lee Theory to Study a
Phase Transition in a Non-Equilibrium System}
\author{Farhad H Jafarpour \footnote{e-mail:farhad@sadaf.basu.ac.ir}\\
Bu-Ali Sina University, Physics Department, Hamadan, Iran}
\maketitle
\begin{abstract}
We study a phase transition in a non-equilibrium model first
introduced in \cite{farhad}, using the Yang-Lee description of
equilibrium phase transitions in terms of both canonical and grand
canonical partition function zeros. The model consists of two
different classes of particles hopping in opposite directions on a
ring. On the complex plane of the diffusion rate we find two
regions of analyticity for the canonical partition function of
this model which can be identified by two different phases. The
exact expressions for both distribution of the canonical partition
function zeros and their density are obtained in the thermodynamic
limit. The fact that the model undergoes a second-order phase
transition at the critical point is confirmed. We have also
obtained the grand canonical partition function zeros of our model
numerically. The similarities between the phase transition in this
model and the Bose-Einstein condensation has also been studied.
\\[5mm]
\end{abstract}

\section{Introduction}
One of the most important activities in the field of equilibrium
statistical physics is the study of the phase transitions;
nevertheless, there is a general framework for the statistical
description of the equilibrium systems and also different
approaches for studying their equilibrium phase transitions. One
of these theories was proposed by Yang and Lee in 1952
\cite{yanglee}. The Yang-Lee theory of equilibrium phase
transitions is based on the zeros of the partition function. It is
known that the zeros of the grand canonical partition function of
finite systems $Z(z)$ (in which $z$ is fugacity) are generally
complex or negative if real. In the thermodynamic limit roots
might move down (or up) and touch the positive real axis. When
this happens, a phase transition occurs because the system can
have different behaviors for $z<z_0$ and $z>z_0$, where $z_0$ is
the value of the root on the real axis. For example the pressure
$P=k_B T \lim_{V\rightarrow\infty}((1/V)\log Z)$ will be
non-analytic and the density $\rho=(\partial / \partial \log
z)(P/k_B T)$ will be discontinuous at the transition point which
predict a first-order phase transition. Similarly, one can
investigate the zeros of the canonical partition function as a
function of complex-temperature and find the same transition
points \cite{gross}. By calculating the line of zeros and also
their density in the thermodynamic limit one can find the
transition point and also the order of transition
exactly.\\
Recently much attention has been focused on one-dimensional out of
equilibrium systems because of their interesting properties such
as first-order phase transitions and spontaneous symmetry breaking
\cite{sz,sch}. However, in contrast to the equilibrium systems
many powerful concepts are missing in this context. For example,
the applicability of the Yang-Lee theory to the non-equilibrium
systems such as one-dimensional driven diffusive models is a quite
non-trivial question and yet without answer. People have tried to
apply the Yang-Lee theory to describe phase transitions in these
models \cite{arndt,evbl}. It seems that one can define similar
quantities such as a grand canonical partition function and then
apply this theory to the out of equilibrium systems without any
problem. In this paper we will apply the Yang-Lee theory to an
exactly solvable one-dimensional non-equilibrium model and
investigate its phase transitions. This model has already been
solved in our previous work and exact results are available
\cite{farhad}. Here we are going to compare our previous results
with those obtained from application of the Yang-Lee theory;
however, we should mention some of the differences between our
work in this paper and what other people have done so far. In
\cite{arndt} the author studies a particle-conserving driven
diffusive model consists of two different classes of particles
with finite densities \cite{ahr} \footnote{This model is known as
AHR model in related literature.} and applies the Yang-Lee theory
to it. He aimed to locate a phase transition induced by varying
the density of particles. By introducing a grand canonical
partition function as a function of a fugacity-like quantity a
first-order phase transition is located by studying the
numerically obtained zeros of this function. In another paper
\cite{evbl} the authors investigate the phase transitions in the
asymmetric simple exclusion process (ASEP) with open boundaries
\cite{gs,dehp} by studying the zeros of the partition function of
the model. In their approach the boundary rates were generalized
to the complex plane. They obtained the distribution of zeros near
the transition point and also their density near the real axis
using the similarities with electrostatic theory. They
could also calculate the line of zeros analytically. \\
The solvability of our model allows us calculate its canonical
partition function exactly. This was done in our previous paper
\cite{farhad} and will be reviewed in the second section. Apart
from numerical estimates for the zeros of the canonical partition
function as a function of reaction rates, we have calculated the
line of zeros and also their density (which determines the order
of the phase transition) analytically using the equilibrium
statistical physics toolbox. Having the exact analytical results
we can discuss the possibility of phase transition and also obtain
its order. Next we will define a grand canonical partition
function and study its zeros in the complex-fugacity plane. The
properties of this function reveal the similarities between the
phase transition in our model and that of a Bose gas. In the last
section we will summarize our results and generalize our approach
to other non-equilibrium models.
\section{The Model}
In \cite{farhad} we introduced a one-dimensional exclusion model
consists of two different classes of particles (we call them
positive and negative particles hereafter) which occupy the sites
of a chain of length $L$ with periodic boundary condition. Each
site of the chain is either empty or occupied by a negative or by
a positive particle. The positive (negative) particles hop to
their immediate right (left) sites with unit rate provided that
the target sites are empty. Adjacent particles with different
charge type might exchange their positions with asymmetric rates
$1$ and $q$. Specifically, the interaction rules are \be
\begin{array}{ccccccc}
\la{rules}
+ & 0   &\longrightarrow& 0 & + & \mbox{with the rate} & 1 \\
0 & -   &\longrightarrow& - & 0 & \mbox{with the rate} & 1 \\
+ & -   &\longrightarrow& - & + & \mbox{with the rate} & q \\
- & +   &\longrightarrow& + & - & \mbox{with the rate} & 1.\\
\end{array}
\ee Assuming that there are $M$ positive particles and only one
negative particle on the chain \footnote{In this case our model is
a special case of AHR model \cite{ahr} in which the number of the
positive and negative particles on the ring are equal.} we showed
that the steady state weights $P({\cal C})$ of the model can be
obtained exactly using a so-called Matrix Product formalism
\cite{dehp}. One can define the sum of these steady state weights
as a quantity which plays the role analogous to the canonical
partition function in equilibrium statistical physics
$$
Z=\sum_{\cal C}P({\cal C}).
$$
It turns out that $Z$ has a closed form in terms of the transition
rate $q$, the number of the positive particles $M$ and the length
of the chain $L$ \be \la{pf} Z_{L,M}(q) = \sum_{i=0}^{M}
\frac{(q-3)(\frac{2}{q})^i+1}{q-2}C_{L-i-2}^{M-i} \ee in which
$C_{i}^{j}=\frac{i!}{j!(i-j)!}$ is the binomial coefficient. In
the thermodynamic limit \be \la{tl} L \ , M \ \longrightarrow
\infty \ \ \mbox{with} \ \rho=\frac{M}{L} \ \mbox{being fixed} \ee
using the steepest decent method it can be shown that \be \la{ab}
\ba{l} \mbox{for} \ \ q<2 \rho \ \ Z_{L,M}(q) \simeq
(\frac{q-3}{q-2})
\frac{(\frac{2}{q})^{L-1}}{(\frac{2}{q}-1)^{L-M-1}}\\
\mbox{for} \ \ q>2 \rho \ \ Z_{L,M}(q) \simeq (1-\rho)(1+\rho
\frac{(q-2\rho)-(q-2)^2}{(q-2)(q-2\rho)})C_{L}^{M}. \ea \ee The
existence of two different phases is apparent. In the same
reference we have shown that the density profile of the positive
particles has an exponential behavior for $q<2\rho$ with a
correlation length $\xi=\vert\ln\frac{q_c}{q}\vert^{-1}$ which
diverges as $q$ approaches its critical value $q_c=2\rho$, while
it is an error function for $q>2\rho$. This proves the existence
of a second-order phase transition from a {\em power-law phase} to
a {\em jammed phase}. Using the Matrix Product formalism one can
also calculate the speed of the different species of particles on
the ring in the thermodynamic limit. Both speeds are linearly
increasing functions of $q$ for $q \le 2 \rho$. However, for $q\ge
2 \rho$ the speed of the positive particles is a constant equal to
$1-\rho$ while the speed of the negative particle is a complicated
increasing function of $q$. In the following section we will apply
the Yang-Lee theory of equilibrium phase transitions to our model.
As we will see, it will not only recover all the above mentioned
results but also shed more light on the unknown aspects of our
problem.
\section{The Partition Function Zeros}
Let us study the phase transition of the our model using the
Yang-Lee theory. We consider the zeros of the canonical partition
function of the model $Z_{L,M}(q)$ given by (\rf{pf}) in the
complex-$q$ plane at fixed $L$ and $M$.
\begin{figure}[htbp]
\setlength{\unitlength}{1mm}
\centering
\includegraphics[height=5cm] {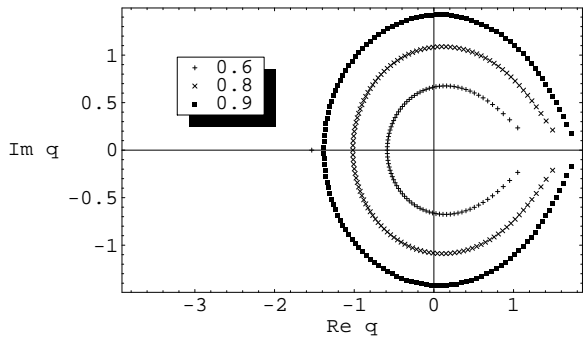}
\caption{The numerical estimates for the roots of $Z_{L,M}(q)$ in
the complex-$q$ plane for different values of density $\rho=0.6,
0.8, 0.9$ and $L=150$. It is seen that the roots accumulate to the
positive real $q$ axis at $1.2$, $1.6$ and $1.8$ respectively.}
\end{figure}
In Figure (1) we have plotted the numerical estimates of these
zeros for a chain of length $L=150$ and three different values of
$\rho$. As can be seen the zeros accumulate slowly to the real $q$
axis at a critical value $q_c=2 \rho$. As we will see later this
accumulation takes place at an angle $\frac{\pi}{4}$ which is the
reminiscent of a second-order phase transition \cite{gross}. In
what follows we will try to find the line of the canonical
partition function zeros and also their density near the positive
real-$q$ axis using the equilibrium statistical physics tools. We
will see that the equilibrium-type calculations give the same
results obtained from the numerical estimates.\\ It has been shown
that the line of zeros can be obtained from \cite{gross} \be
\la{formula1} Re \ g_1=Re \ g_2. \ee In equilibrium statistical
physics $g$ is the extensive part of free energy and is generally
a function of the temperature (here $q$) and the density of
particles $\rho$. The indexes $1$ and $2$ show the values of the
function $g$ in the right and the left hand side of the critical
point. Here we define this function as \be \la{gg}
g(q,\rho)=\lim_{L,M\rightarrow\infty} \frac{1}{L}\ln Z_{L,M}(q)
\ee where $\lim \cdots$ is in fact the thermodynamic limit given
by (\rf{tl}). Using the asymptotic behavior of $Z_{L,M}(q)$ given
by (\rf{ab}) we can calculate $g(q,\rho)$ in each phase. After
further calculations we obtain \be \la{g} \ba{l}
\mbox{for} \ \ q<2 \rho \ \ g(q,\rho)=\ln \frac{2}{ q^{\rho}(2-q)^{1-\rho}}\\
\mbox{for} \ \ q>2 \rho \ \ g(q,\rho)=\ln
\frac{1}{\rho^\rho(1-\rho)^{1-\rho}}. \ea \ee By substituting
(\rf{g}) in (\rf{formula1}) we find the following equation for the
line of zeros in the complex-$q$ plane \be \la{lz}
((\frac{2-x}{1-\rho})^2+
(\frac{y}{1-\rho})^2)^{1-\rho}((\frac{x}{\rho})^2+(\frac{y}{\rho})^2)^{\rho}=4.
\ee in which $ x \equiv Re(q)$ and $y \equiv Im(q)$. In Figure (2)
we have plotted both (\rf{lz}) and the numerical estimates of
zeros given in the Figure (1) for three values of $\rho$.
\begin{figure}[htbp]
\setlength{\unitlength}{1mm}
\begin{picture}(0,0)
\put(-1,26){\makebox{$\scriptstyle y$}}
\put(30,-1){\makebox{$\scriptstyle x$}}
\end{picture}
\centering
\includegraphics[height=5cm] {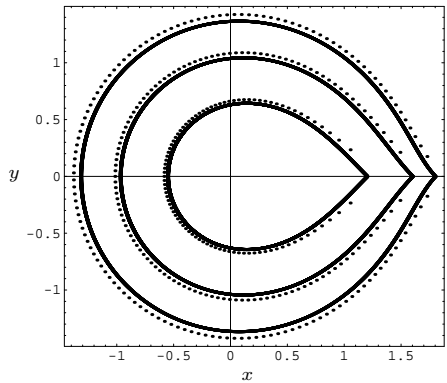}
\caption{Plot of the equation (\rf{lz}) (solid lines) for three
values of density of the positive particles $\rho=0.6$ (inner),
$\rho=0.8$ (center) and $\rho=0.9$ (outer). They cross the
positive $x$ axis at $x=2\rho$. The dotted lines belong to the
numerical estimates of the canonical partition function zeros for
the same values of densities and are taken from the Figure (1).}
\end{figure}
As can be seen the curves lie on the numerical estimates of the
canonical partition function zeros. The small difference belongs
to the fact that the numerical estimates have not been calculated
in real thermodynamic limit. The curves also cross the positive
$x$ axis at $x_c=2\rho$ which is the transition point as we had
mentioned above. In the equilibrium Yang-Lee theory it is well
known that the density of zeros on the real positive axis is zero
at a second-order phase transition. In order to obtain the density
of zeros $\mu$ in this region (on the positive real $x$ axis and
near the critical point) we use the following equation first
introduced in \cite{gross} \be \la{formula2} 2\pi
\mu(s,\rho)=\frac{\partial}{\partial s} Im(g_1-g_2) \ee in which
$s$ is the arc length of the line of the zeros which is zero at
the critical point and increases along with the line of zeros in
the positive $y$ direction. The values of $g_1$ and $g_2$ are
given by (\rf{g}). In order to calculate $\mu(s,\rho)$ first we
find an equation for the line of zeros which is valid for small
$y$'s in the vicinity of the critical point $x_c=2 \rho$.
\begin{figure}[htbp]
\setlength{\unitlength}{1mm}
\centering
\includegraphics[height=5cm] {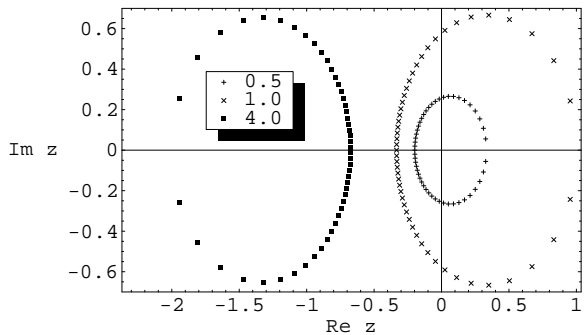}
\caption{Plot of the numerical estimates of the zeros of the grand
canonical partition function (\rf{gcpf}) for $q=0.5$, $1.0$ and
$4.0$. The length of the chain is $L=50$.}
\end{figure}
Using (\rf{lz}) it can be shown that this is actually a line \be
\la{line} y=2 \rho-x \ \ \mbox{for} \ \ \vert x-2\rho\vert \ll 1 \
, \  0<y \ll 1. \ee The equation (\rf{line}) confirms that the
accumulation of zeros in the vicinity of the real-$q$ axis takes
place at angle $\frac{\pi}{4}$. Using (\rf{g}), (\rf{formula2})
and (\rf{line}) we find for $\rho\neq 1$ \be
2\pi\mu(y,\rho)=-\frac{\partial y}{\partial
s}\frac{\partial}{\partial y} Im(\ln \frac{2}{
(x+iy)^{\rho}(2-x-iy)^{1-\rho}}) \propto \frac{y}{2\rho(1-\rho)}.
\ee Therefore, as $y$ approaches to zero the density of zeros
becomes zero as we expect for a second-order phase transition.
Comparing the numerical data given in Figure (1) with the results
obtained from the application of the equilibrium statistical
physics tools shows good agreement between the two approaches;
therefore, we it is reasonable to believe that the analytical
approach presented here gives the exact result.\\
It is also interesting to look for the zeros of the grand
canonical partition function of our model. In this case we can
investigate the similarities between the phase transition in our
model with that of a Bose gas. As opposed to the equilibrium
statistical physics, the definition of a grand canonical ensemble
for the steady state of a non-equilibrium system is not unique. We
adopt the following definition first used in \cite{djls} \be
\la{gcpf} Z_{L}(q,z)=\sum_{M=0}^{L-1}z^M Z_{L,M}(q) \ee in which
$z$ is the fugacity of the positive particles and $Z_{L,M}(q)$ is
given by (\rf{pf}).
\begin{figure}[htbp]
\setlength{\unitlength}{1mm}
\begin{picture}(0,0)
\put(-1,30){\makebox{$\scriptstyle \rho(z)$}}
\put(42,1){\makebox{$\scriptstyle z$}}
\end{picture}
\centering
\includegraphics[height=5cm] {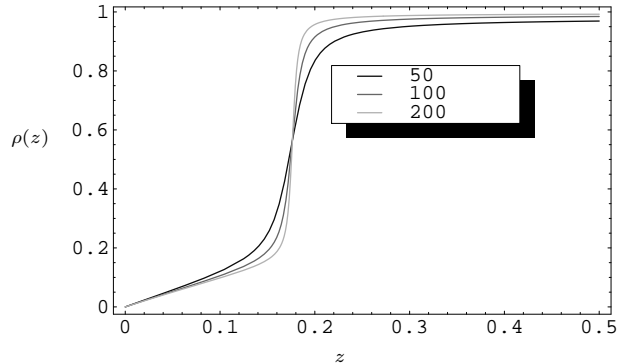}
\caption{Plot of the density of the positive particles (\rf{den})
as a function of fugacity $z$ for $L=50,100$ and $200$ at $q=0.3$.
In the thermodynamic limit there is a finite discontinuity in the
density of particles.}
\end{figure}
It is seen that the grand canonical partition function (\rf{gcpf})
is a polynomial of degree $L-1$ in $z$ so, in the complex-$z$
plane it has $L-1$ zeros $z_i$ and can be written as $$
Z_{L}(q,z)=\prod_{i=1}^{L-1}(z-z_i).$$ Using (\rf{pf}) one can
easily calculate the grand canonical partition function
(\rf{gcpf}) explicitly \be \la{gcpf2}
Z_L(q,z)=(\frac{q-3}{q-2})[(\frac{2z}{q})^{L-1}+\frac{(1+z)^{L-1}}{(q-3)}+
\frac{(\frac{2z}{q})^{L-1}-(1+z)^{L-1}}{(\frac{2z}{q})-(1+z)}].\ee
The fugacity of the positive particles in (\rf{gcpf}) has to be
fixed by density of them \be \la{den}
\rho(z)=\frac{z}{L}\frac{\partial}{\partial z} \ln Z_{L}(q,z). \ee
Let us examine the zeros of the grand canonical partition function
given by (\rf{gcpf2}) in the complex-$z$ plane. Since the grand
canonical partition function is a real polynomial with positive
coefficients (they are sum over probabilities) the zeros come in
complex conjugate pairs and the real roots are negative. In Figure
(3) we have plotted the numerical estimates of the zeros of
(\rf{gcpf2}) in complex-$z$ plane for three values of $q$. As long
as $q<2$ the roots lie on a vertical elliptic. It also appears
that the zeros approach the real $z$ axis at an angle $\pi \over
2$. It is a sign of a first-order phase transition. For $q>2$ all
of the roots have negative real parts; therefore, we do not expect
any phase transition to take place. For $q=2$, we should take the
limit of (\rf{pf}) since it is undefined at this point. In the
following we will investigate our model in order to see its
similarities with the Bose-Einstein condensation. Using
(\rf{gcpf2}) one can calculate (\rf{den}) in each phase in the
thermodynamic limit explicitly. It turns out that \be \la{den2}
\rho(z)= \left\{
\begin{array}{ll}
\frac{z}{z+1}& \mbox{for } z<\frac{q}{2-q} \\
1 & \mbox{for } z>\frac{q}{2-q}.
\end{array}
\right. \ee In Figure (4) we have plotted (\rf{den}) for different
values of $L$. As can be seen the $L$ dependence of the density
suggests that in the thermodynamic limit, $\rho$ increases with
$z$ and then at a specific point $z_0=\frac{q}{2-q}$ it has a
finite discontinuity. At this point, $z$ does not fix the density
anymore and the system undergoes a first-order phase transition.
\begin{figure}[htbp]
\setlength{\unitlength}{1mm}
\begin{picture}(0,0)
\put(-5,30){\makebox{$\scriptstyle P$}}
\put(42,-2){\makebox{$\scriptstyle \frac{1}{\rho}$}}
\end{picture}
\centering
\includegraphics[height=5cm] {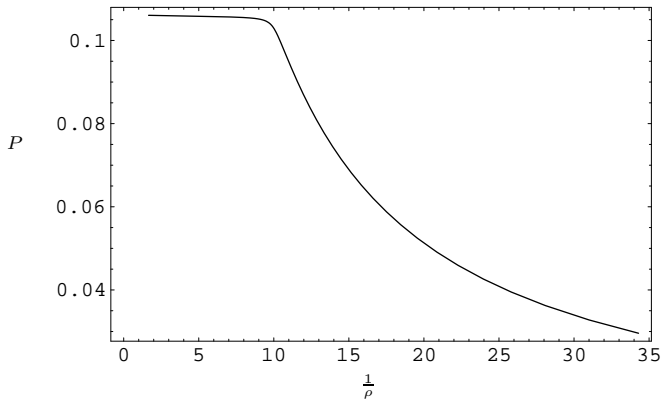}
\caption{ The pressure $P$ as a function of $\rho^{-1}$ for
$q=0.2$ and $L=15000$.}
\end{figure}
The finite jump in the density is related to the finite density of
roots of (\rf{gcpf2}), $\mu(z)$, at the real $z$ axis \be
\la{denroot} \mu(z_0)=\frac{1-\rho(z_0)}{2\pi z_0}. \ee The
critical fugacity $z_0$ can also be obtained by extrapolating the
real part of the nearest root\footnote{In fact there are two of
them since the roots appear in complex conjugate pairs.} to the
real positive $z$ axis for $L\rightarrow \infty$. In the
Bose-Einstein condensation the density of the particles has such a
behavior where the conservation of the number of particles is
broken \cite{pathria}. Another interesting quantity is the
pressure which can be defined analogously to equilibrium physics
\be \la{pre} P(z)=\frac{1}{L}\ln Z_{L}(q,z). \ee However, $P$ is
not the physical pressure of the particles in the current context.
The particles pressure in each phase in the thermodynamic limit
can be calculated using (\rf{gcpf2}) \be \la{pre2} P(z)= \left\{
\begin{array}{ll}
\ln(1+z)& \mbox{for } z<\frac{q}{2-q} \\
\ln(\frac{2z}{q}) & \mbox{for } z>\frac{q}{2-q}.
\end{array}
\right. \ee Now by using (\rf{den2}) and (\rf{pre2}) it can easily
be verified that in the power-law phase ($\rho<\frac{q}{2}$) the
particles pressure $P$ as a function of density has the form
$P(\rho)=\ln(\frac{1}{1-\rho})$ while in the jammed phase
($\rho>\frac{q}{2}$), as can be seen from (\rf{den2}), the
density-fugacity relation (\rf{den}) breaks down and results in
$\rho=1$\footnote{Physically, this is related to the existence of
a shock in this phase.}; however, the particles pressure remains
constant in this phase $P(\rho)= \ln(\frac{q}{2-q})$. In Figure
(5) we have plotted (\rf{pre}) as a function of the inverse of
density of particles $\rho^{-1}$ for $L=15000$ and $q=0.2$. The
critical density in this case is $\rho=0.1$. As can be seen for
$\frac{1}{\rho}>10$ the pressure decreases as the density gets
smaller; however, for $\frac{1}{\rho}<10$ it is nearly constant.
This {\em isotherm} (here $q$ instead of temperature $T$) is
similar to the isotherm of the free Bose gas when the
Bose-Einstein condensation takes place. One can also look at the
compressibility $\kappa$ which is defined as \be \la{comp}
\kappa=L\zeta^2 \ee in which \be \la{zeta}
\zeta=\frac{\sqrt{\langle\rho^2\rangle-{\langle \rho
\rangle}^2}}{\rho} \ee and the fluctuation of the density of the
positive particles can be obtained using \be \la{fluc}
\langle\rho^2\rangle-{\langle \rho
\rangle}^2=\frac{z}{L}\frac{\partial}{\partial z}
(\frac{z}{L}\frac{\partial}{\partial z} \ln Z_{L}(q,z)). \ee
Figure (6) shows $\zeta$ as a function of the chain length $L$ for
$\rho=0.6$ and two values of $q$ above and bellow of the
transition point.
\begin{figure}[htbp]
\setlength{\unitlength}{1mm}
\begin{picture}(0,0)
\put(-1,30){\makebox{$\scriptstyle \zeta$}}
\put(42,1){\makebox{$\scriptstyle L$}}
\end{picture}
\centering
\includegraphics[height=5cm] {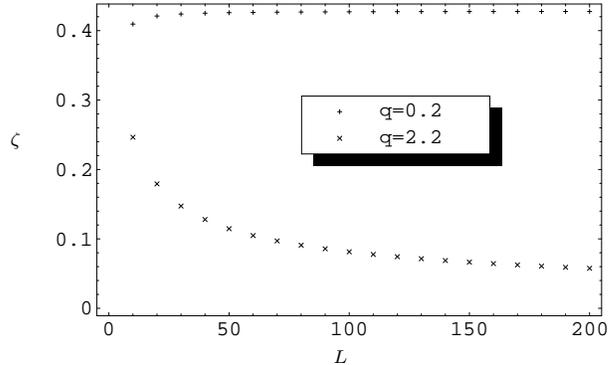}
\caption{ Plot of (\rf{zeta}) as a function of $L$ for $\rho=0.6$
and two values of $q$. }
\end{figure}
The transition point in this case occurs at $q_c=1.2$. It can be
seen that $\zeta$ is a decreasing function of $L$ in the power-law
phase ($q>q_c$) while it is nearly constant in the jammed phase
($q<q_c$). Using (\rf{gcpf2}) and (\rf{fluc}) it can be verified
that in the thermodynamic limit $\zeta$ drops down as
$\frac{1}{\sqrt{L}}$ in the power-law phase while in the jammed
phase it remains constant which gives a divergent compressibility.
\section{Concluding Remarks}
In this paper we have used the Yang-Lee description of equilibrium
phase transitions in terms of the zeros of both canonical and
grand canonical partition function to study a non-equilibrium
phase transition in a driven diffusive system. By studying the
canonical partition function zeros a second-order phase transition
was predicted. This is in quite close agreement with our previous
results in \cite{farhad}. The line of zeros and also their density
on the real axis near the transition point were obtained exactly.
By introducing the grand canonical partition function of our model
(\rf{gcpf}) the similarities between the phase transition in our
model and the one seen in the equilibrium Bose gas were elegantly
observed.\\
The approach that we used in this paper can also be applied to
other models. In a similar model consists of a group of first
class particles hopping behind a slow (or a second class) particle
on a closed chain of length $L$, with the following interaction
rules \be
\begin{array}{ccccccc} \la{rules2}
1 & 0   &\longrightarrow& 0 & 1 & \mbox{with the rate} & 1 \\
2 & 0   &\longrightarrow& 0 & 2 & \mbox{with the rate} & \alpha.
\end{array}\ee
it is shown that by choosing the right reaction rates the
probability distribution for the stationary state can be mapped to
the one obtained for an ideal Bose gas \cite{ev}. The
thermodynamic limit of the canonical partition function of this
model is obtained in \cite{ev} \be \la{ab2} \ba{l} \mbox{for} \ \
\alpha<1-\rho \ \ Z_{L,M}(\alpha)
\simeq \frac{(1-\alpha)^{1-M}}{(\alpha)^{L-M}}\\
\mbox{for} \ \ \alpha>1-\rho \ \ Z_{L,M}(\alpha) \simeq
\frac{\alpha\rho^2}{\rho+\alpha-1}C_{L}^{M} \ea \ee in which
$\rho=\frac{M}{L}$ is the density of the first class particles.
Now using (\rf{formula1}) and (\rf{gg}) we find the line of zeros
in complex-$\alpha$ plane \be
((\frac{1-x}{\rho})^2+(\frac{y}{\rho})^2)^{\rho}
((\frac{x}{1-\rho})^2+(\frac{y}{1-\rho})^2)^{1-\rho}=1\ee where $x
\equiv Re(\alpha)$ and $y \equiv Im(\alpha)$. This function
crosses the real positive $\alpha$ axis at $\alpha_c=1-\rho$ at an
angle $\frac{\pi}{4}$. It can be seen that the thermodynamic limit
of the canonical partition function of our model (\rf{ab}) is
quite similar to (\rf{ab2}). Using (\rf{formula1}) one can also
obtain the line of zeros for the ASEP with open boundaries
exactly.
\begin{figure}[htbp]
\setlength{\unitlength}{1mm}
\begin{picture}(0,0)
\put(-1,26){\makebox{$\scriptstyle y$}}
\put(42,-2){\makebox{$\scriptstyle x$}}
\put(61,35){\makebox{$\scriptstyle \beta=1$}}
\put(46,30){\makebox{$\scriptstyle \beta=\frac{1}{3}$}}
\end{picture}
\centering
\includegraphics[height=5cm] {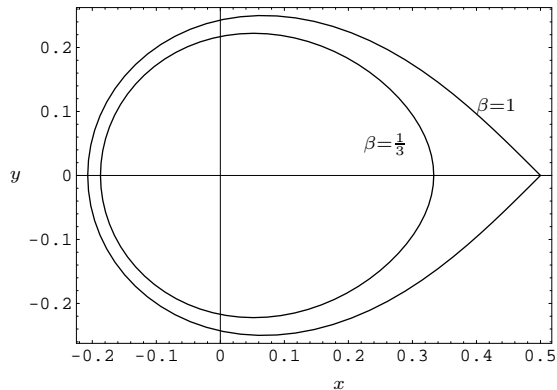}
\caption{Plot of the equation (\rf{asep}) for $\beta=\frac{1}{3}$
and $\beta=1$.}\end{figure} For this model with two parameters
$\alpha$ and $\beta$ as injection and extraction rates of
particles, the line of zeros in the complex-$\alpha$ plane (and
$\beta$ being fixed) is \be \la{asep} (-x^2+x+y^2)^2+(y-2xy)^2=J^2
\ee in which we have defined  $x \equiv Re(\alpha)$, $y \equiv
Im(\alpha)$ and the current of particles is
$$ J= \left\{
\begin{array}{ll}
\frac{1}{4}   & \mbox{ if }  \beta \geq \frac{1}{2} \\ \\
\beta(1-\beta)   & \mbox{ if } \beta \leq \frac{1}{2}.
\end{array}
\right. $$ The investigation of (\rf{asep}) confirms the existence
of two phase transitions in the system. In Figure (7) we have
plotted the line of zeros (\rf{asep}) for two values of $\beta$.
It can be seen that (\rf{asep}) gives exactly the same results
obtained in \cite{evbl} from the numerical estimates of the
canonical partition function zeros. In \cite{evbl} the line of
zeros is stated to be $\vert \alpha(1-\alpha)\vert=J$. By
putting $\alpha=x+iy$ one finds (\rf{asep}) for the line of zeros.\\
Apart from the above mentioned models one can show that our
approach can be applied to many other non-equilibrium systems like
those introduced in \cite{others}. Work in this direction is in
progress \cite{fnew}.
\\[5mm]
{\large {\bf Acknowledgements}}
\\[5mm]
I would like to thank G. M. Sch\"utz for reading the manuscript
and his comments and also the {\em Max-Planck Institute f\"ur
Physik Komplexer Systeme} where the preliminary calculations for
this work were done.
\\[1cm]
{\Large {\bf References}}

\end{document}